\documentclass[11pt,superscriptaddress,aps,prd,preprint]{revtex4}

\everymath{\displaystyle}
\usepackage{graphicx}

\usepackage[T1]{fontenc}
\usepackage{amsmath}
\usepackage{amssymb}
\usepackage{graphicx}
\usepackage{slashed}
\usepackage{xcolor}

\newcommand{\bea}{\begin{eqnarray}}
\newcommand{\eea}{\end{eqnarray}}

\begin{document}

\title{Bhabha scattering in Very Special Relativity at finite temperature}

\author{A. F. Santos}
\email{alesandroferreira@fisica.ufmt.br}
\affiliation{Instituto de F\'{\i}sica, Universidade Federal de Mato Grosso,\\
78060-900, Cuiab\'{a}, Mato Grosso, Brazil}

\author{Faqir C. Khanna \footnote{Professor Emeritus - Physics Department, Theoretical Physics Institute, University of Alberta\\
Edmonton, Alberta, Canada}}
\email{khannaf@uvic.ca; fkhanna@ualberta.ca}
\affiliation{Department of Physics and Astronomy, University of
Victoria,BC V8P 5C2, Canada}

\begin{abstract}

In this paper the differential cross section for Bhabha scattering in the Very Special Relativity (VSR) framework is calculated. The main characteristic of the VSR is to modify the gauge invariance. This leads to different types of interactions appearing in a non-local form. In addition, using the Thermo Field Dynamics formalism, thermal corrections for the differential cross section of Bhabha scattering in VSR framework are obtained.

\end{abstract}

\maketitle

\section{Introduction}

Symmetry underlies all theories that describe nature. The fundamental laws of nature are invariant under the Poincaré group, that is the set of Lorentz transformations plus space-time translations. Although Lorentz symmetry has been tested experimentally to a high degree of precision, possible tiny violations of Lorentz symmetry may emerge in theories that attempt to unify all known forces. Opportunities to detect experimentally these violations will likely arise at the Planck scale, $\sim 10^{19}\, \mathrm{GeV}$. Minimal modifications of standard model or alternative models have been proposed in order to attempt to
understand possible Lorentz violation. One attempt is the Standard Model Extension (SME) \cite{SME1, SME2} which contains the Standard Model, General Relativity and all possible operators that break Lorentz symmetry. Another interesting way is the so called Very Special Relativity (VSR) \cite{Cohen1, Cohen2}.

The main characteristic of VSR is that the laws of nature are not invariant under the whole Lorentz group but instead are invariant under subgroups of the
Lorentz group that still preserves the basic elements of special relativity like the constancy of the velocity of light. In the VSR framework a modified gauge symmetry is present, then a variety of new gauge invariant interactions are permitted. The most interesting of these subgroups of VSR are the $SIM(2)$ and $HOM(2)$. The so-called Homothety group $HOM(2)$ has three parameters and is generated by $T_1=K_x+J_y$, $T_2=K_y-J_x$ and $K_z$, wiht $\vec{J}$ and $\vec{K}$ being the generators of rotations and boosts, respectively. The $SIM(2)$, called the Similitude group consists of $HOM(2)$ group plus the $J_z$ generator. These subgroups do not have invariant tensor fields. These subgroups preserve the direction of a light-like four-vector $n_\mu$. Then, theories that are invariant under these subgroups have a preferred direction in Minkowski space-time. In addition, all local operators preserving $HOM(2)$ or $SIM(2)$ also preserve Lorentz symmetry. However, non-local terms that violate Lorentz symmetry can be constructed as ratios of contractions of $n_\mu$ with other kinematic vectors. These non-local terms that violate Lorentz symmetry are invariant under $HOM(2)$ or $SIM(2)$.

Several applications in the VSR framework have been considered. For example, implications for neutrino physics of VSR are explored. The generation of a neutrino mass without lepton number violation nor sterile neutrinos has been admitted. In the ultra-relativistic limit, VSR and conventional neutrino masses are indistinguishable. However, VSR effects are significant near the beta decay endpoint where neutrinos are not ultra-relativistic \cite{Cohen2}. A supersymmetric field theory that is translation and $SIM(2)$ invariant, but not Lorentz invariant has been formulated. In this theory the number of supersymmetries are half that required of a standard Lorentz invariant theory. This leads to a modified SUSY algebra \cite{Cohen3}. The $SIM(2)$ superspace formulation of the supersymme\-tric Yang-Mills gauge theory minimally coupled to chiral superfields has been discussed. Then the super-Poincare invariant supersymmetric Yang-Mills theory to $SIM(2)$ superspace formalism has been formulated \cite{Vo}. Noncommutative implications of VSR are investigated. It has been established that the light-like Moyal noncommutative space provides a consistent framework for $T(2)$ group of VSR. The other VSR subgroups are discarded, if the origin of Lorentz violation is in the noncommutative structure of space-time, since the corresponding noncommutative spaces are not translationally invariant \cite{MM}. A deformation of the subgroup $SIM(2)$, known as $DSIM(2)$, has been introduced. A novel non-commutative spacetime structure, underlying the $DSIM(2)$ has been presented. It allows us to construct explicitly the generators of the group \cite{Das}. Non-abelian fields in VSR have been analyzed. A covariant derivative and modified gauge transformations are defined. Actions for matter fields coupled to the VSR gauge fields have been constructed. The BRST formalism is used to obtain the propagators and vertices of pure VSR Yang–Mills theory. In addition, the non-abelian theory in VSR is renormalizable and asymptotically free \cite{Riv}. A VSR inspired modification of Maxwell-Chern-Simons electrodynamics has been constructed. The classical dynamics for this model is analyzed. The solution for the electric field and static energy for this configuration is obtained. The interaction energy between opposite charges and a finite expression for the static potential is derived \cite{Bufalo}. The thermodynamical properties of the quantum electrodynamics in the VSR framework have been studied. The thermal effects are introduced using the Matsubara imaginary-time formalism. To explore new interactions of VSR, the effective Lagrangian at one- and two-loop order in VSR are calculated \cite{Bufalo2}. Consequences of a gauge invariant photon mass in VSR are discussed. In this context, the Maxwell-VSR equations and the modified Feynman rules are available. The Coulomb scattering and radiative corrections have been computed \cite{Soto}.  One loop quantum corrections to the photon self energy, electron self energy and vertex in the electrodynamics sector of VSR are calculated. An appropriate regulator, based on the calculation of integrals using the Mandelstam–Leibbrandt prescription has been introduced \cite{Alf}. The differential cross section for Bhabha and Compton scattering for the quantum electrodynamics defined in the framework of SIM(2) of VSR have been calculated \cite{Bufalo3}. Although there are many investigations in the presence of VSR non-local terms, corrections due to finite temperature are still missing. The main objective of this paper is to calculate the differential cross section for Bhabha scattering in VSR at finite temperature. This scattering process involves electrons and positrons with a photon as an intermediate particle. The Thermo Field Dynamics (TFD) formalism is used to introduce finite temperature.

TFD is an approach to introduce temperature effects in quantum field theory \cite{Umezawa1, Umezawa2, Umezawa22, Khanna1, Khanna2}. It is known as real time formalism. Its basic elements are: (i) the doubling of the original Hilbert space, which consists of Hilbert space composed of the original, $S$ and a tilde space $\tilde{S}$ (dual space). These two spaces are mapped by the tilde (or dual) conjugation rule. (ii) The Bogoliubov transformation, that is a rotation involving these two spaces. As a consequence the propagator is written in two parts: $T = 0$ and $T\neq 0$ components. Another important feature of the TFD formalism is that it preserves the time-evolution once the temperature is identified with a rotation in a duplicated Hilbert space.

This paper is organized as follows. In section II, QED in Very Special Relativity is presented. Some attention to the new interaction between fermions and photon is given. In section III, the TFD formalism is introduced. The photon propagator at finite temperature is discussed. In section IV, the differential cross section for Bhabha scattering in VSR at finite temperature is calculated. In section V, some concluding remarks are presented.

\section{QED in Very Special Relativity}

In this section the QED Lagrangian in VSR is presented. This Lagrangian describes the interaction between fermions and photon. The $SIM(2)$ VSR-invariant Lagrangian is given as
\bea
{\cal L}=-\frac{1}{4}\tilde{F}_{\mu\nu}\tilde{F}^{\mu\nu}+\bar{\psi}\left(i\gamma^\mu \nabla_\mu-m_e\right)\psi,
\eea
where $m_e$ is the electron mass. Here, the field strength is defined in terms of the wiggled derivative as
\bea
\tilde{F}_{\mu\nu}=\tilde{\partial}_\mu A_\nu-\tilde{\partial}_\nu A_\mu
\eea
with 
\bea
\tilde{\partial}_\mu=\partial_\mu+\frac{m^2}{2}\frac{n_\mu}{n\cdot \partial}
\eea
being the wiggle derivative. The $m$ parameter sets the scale for the VSR effects, and $n_\mu$ is a light-like four-vector that represents the preferred null direction given as $n_\mu=(1,0,0,1)$. It is important to note that, using the wiggle derivative in the field strength the VSR Maxwell equation $\tilde{\partial}_\mu \tilde{F}^{\mu\nu}=0$ shows that each component of the gauge field satisfies a Klein-Gordon equation, then a massive gauge field is developed \cite{Riv, Lee}. In this new gauge structure the minimal coupling among fermions and photons is determined by a new covariant derivative given by
\bea
\nabla_\mu\psi=D_\mu\psi+\frac{1}{2}\frac{m^2}{n\cdot D}n_\mu\psi,
\eea
where the usual covariant derivative is defined as $D_\mu=\partial_\mu-ieA_\mu$. It is important to note that, due to the non-local character of term $1/(n\cdot D)$ in the covariant derivative, there is an infinite number of interactions in the coupling $e$. The Feynman rules for these interactions are constructed using the Wilson lines approach. For details see \cite{Dunn}. Then part of the Lagrangian that describes just the interaction between fermions and photons is
\bea
{\cal L}_I=-e\bar{\psi}\gamma^\mu\psi A_\mu-e\bar{\psi}\frac{m^2}{2}\frac{\slashed{n}n^\mu}{(n\cdot p_1)(n\cdot p_2)}\psi A_\mu.
\eea
The first term describes the usual QED interaction and the second term is a new interaction, due to the VSR characteristics that leads to violation of the Lorentz symmetries. The main objective is to determine VSR modifications for Bhabha scattering at the tree level. Then it is sufficient to obtain the Feynman rules only for the vertex $\left\langle\bar{\psi}\psi A\right\rangle$ \cite{Dunn}. This vertex has the form
\bea
{\cal V}^\mu(p_1,p_2)= -ie\left[\gamma^\mu+\frac{m^2}{2}\frac{\slashed{n}\,n^\mu}{(n\cdot p_1)(n\cdot p_2)}\right].\label{vertex}
\eea
The photon propagator is given as
\bea
iD_{\mu\nu}(\tilde{p})=\frac{\eta_{\mu\nu}}{\tilde{p}^2},
\eea
where the gauge propagator has a massive pole $\tilde{p}^2=p^2-m^2$. The main objective is to calculate the differential cross section for a scattering process of VSR-QED at finite temperature, in the next section the TFD formalism is introduced.

\section{Introduction to TFD formalism}

TFD is a thermal formalism whose main characteristic is to show that the thermal average of any operator {\cal D} is equal to its temperature dependent vacuum expectation value, i.e., $\langle D \rangle=\langle 0(\beta)| D|0(\beta) \rangle$. Here $|0(\beta) \rangle$ is a thermal vacuum, where $\beta\propto\frac{1}{T}$, with $T$ begin the temperature. In order to satisfy this requirement the doubling of the Hilbert space is considered and the temperature effects are introduced by the Bogoliubov transformation. The doubling of Hilbert space is given by the tilde ($^\thicksim$) conjugation rules, such that the expanded space is $S_T=S\otimes \tilde{S}$, with $S$ being the standard Hilbert space and $\tilde{S}$ the dual (tilde) space. In this formalism each operator in $S$ is associated with two operators in $S_T$. For example, if $o$ is an operator in $S$, then in $S_T$ we have two operators associated to $o$, i.e., $O=o\otimes 1$ and $\tilde{O}=1\otimes o$.

The Bogoliubov transformation consists in a rotation in the tilde and nontilde variables. For fermions with $c_p^\dagger$ and $c_p$ being creation and annihilation operators respectively, Bogoliubov transformations are
\bea
c_p&=&\mathsf{u}(\beta) c_p(\beta) +\mathsf{v}(\beta) \tilde{c}_p^{\dagger }(\beta), \label{f1}\\
c_p^\dagger&=&\mathsf{u}(\beta)c_p^\dagger(\beta)+\mathsf{v}(\beta) \tilde{c}_p(\beta),\label{f2}\\
\tilde{c}_p&=&\mathsf{u}(\beta) \tilde{c}_p(\beta) -\mathsf{v}(\beta) c_p^{\dagger}(\beta),\label{f3} \\
\tilde{c}_p^\dagger&=&\mathsf{u}(\beta)\tilde{c}_p^\dagger(\beta)-\mathsf{v}(\beta)c_p(\beta),\label{f4}
\eea
where $\mathsf{u}(\beta) =\cos \theta(\beta)$ and $\mathsf{v}(\beta) =\sin \theta(\beta)$. The anti-commutation relations for creation and annihilation operators  are similar to those at zero temperature
\bea
\left\{c(k, \beta), c^\dagger(p, \beta)\right\}&=&\delta^3(k-p),\nonumber\\
 \left\{\tilde{c}(k, \beta), \tilde{c}^\dagger(p, \beta)\right\}&=&\delta^3(k-p),\label{ComF}
\eea
and other anti-commutation relations are null.

For bosons with $a_p^\dagger$ and $a_p$ being creation and annihilation operators respectively, the Bogoliubov transformations are
\bea
a_p&=&\mathsf{u}'(\beta) a_p(\beta) +\mathsf{v}'(\beta) \tilde{a}_p^{\dagger }(\beta), \\
a_p^\dagger&=&\mathsf{u}'(\beta)a_p^\dagger(\beta)+\mathsf{v}'(\beta) \tilde{a}_p(\beta),\\
\tilde{a}_p&=&\mathsf{u}'(\beta) \tilde{a}_p(\beta) +\mathsf{v}'(\beta) a_p^{\dagger}(\beta), \\
\tilde{a}_p^\dagger&=&\mathsf{u}'(\beta)\tilde{a}_p^\dagger(\beta)+\mathsf{v}'(\beta)a_p(\beta),
\eea
where $\mathsf{u}'(\beta) =\cosh \theta(\beta)$ and $\mathsf{v}'(\beta) =\sinh \theta(\beta)$. Algebraic rules for thermal operators are
\bea
\left[a(k, \beta), a^\dagger(p, \beta)\right]&=&\delta^3(k-p),\nonumber\\
 \left[\tilde{a}(k, \beta), \tilde{a}^\dagger(p, \beta)\right]&=&\delta^3(k-p),\label{ComB}
\eea
and other commutation relations are null.

It is important to note that, in this formalism the propagator of any particle is written in two parts: one describes the flat space-time contribution and the other displays the thermal and/or the topological effect. As an example, let's write the photon propagator at finite temperature which is used to calculate the transition amplitude of the Bhabha scattering.

The photon propagator is defined as
\bea
iD_{\mu\nu}(x-y)&=&\langle 0(\beta)|\mathbb{T}A_\mu(x)A_\nu(y)|0(\beta)\rangle\nonumber\\
&=&\theta(t_x-t_y)\langle 0(\beta)|A_{\mu}(x)A_{\nu}(y)|0(\beta)\rangle + \theta(t_y-t_x)\langle 0(\beta)|A_{\nu}(y)A_{\mu}(x)|0(\beta)\rangle\nonumber,
\eea
where $\mathbb{T}$ is the time ordering operator, $\theta(t_x-t_y)$ is the step function and $A_\mu(x)$, the free field solution, is given by
\bea
A_{\mu}(x)=\int\frac{d^3p}{\sqrt{2\omega_p(2\pi)^3}}\sum_\lambda\epsilon_{\mu}(p,\lambda)\left(a_{p,\lambda}e^{-ip_\rho x^\rho}+a_{p,\lambda}^\dagger e^{ip_\rho x^\rho}\right),\label{A1}
\eea
with $\epsilon_{\mu}(p,\lambda)$ being the polarization vector and $\omega_p=\sqrt{p^2+m^2}$ is the photon dispersion relation.

The two point function in TFD is a thermal doublet, and has $2\times2$ matrix structure
\bea
\left( \begin{array}{c} A^1_{\mu} \\ A^2_{\mu} \end{array} \right)=\left( \begin{array}{c} A_{\mu} \\ {\tilde A_{\mu}^\dagger} \end{array} \right).\label{doublet1}
\eea
Then the photon propagator becomes 
\bea
iD^{ab}_{\mu\nu}(x-y)&=&\theta(t_x-t_y)\langle 0(\beta)|A^a_{\mu}(x)A^b_{\nu}(y)|0(\beta)\rangle\nonumber\\
& +& \theta(t_y-t_x)\langle 0(\beta)|A^b_{\nu}(y)A^a_{\mu}(x)|0(\beta)\rangle,\label{Delta}
\eea
where $a,b=1,2$ and $\mu,\,\nu$ are tensor indices. Using that $\sum_{\lambda}\epsilon_{\mu}(p,\lambda)\epsilon_{\nu}(p,\lambda)=\eta_{\mu\nu}$ and after some calculations, the photon propagator at finite temperature is 
\bea
D_{\mu\nu}(\tilde{p},\beta)=D_{\mu\nu}^{(0)}(\tilde{p})+D_{\mu\nu}^{(\beta)}(\tilde{p}),\label{photon1}
\eea
where $D_{\mu\nu}^{(0)}(\tilde{p})$ and $D_{\mu\nu}^{(\beta)}(\tilde{p})$ are zero and finite temperature parts respectively. Explicitly 
\bea
D_{\mu\nu}^{(0)}(\tilde{p})&=&\frac{\eta_{\mu\nu}}{\tilde{p}^2}\tau,\nonumber\\
D_{\mu\nu}^{(\beta)}(\tilde{p})&=&-\frac{2\pi i\delta(\tilde{p}^2)}{e^{\beta \tilde{p}_0}-1}\left( \begin{array}{cc}1&e^{\beta \tilde{p}_0/2}\\e^{\beta \tilde{p}_0/2}&1\end{array} \right)\eta_{\mu\nu},
\eea
where $\tau=\left( \begin{array}{cc}1 & 0 \\ 0 & -1\end{array} \right)$. More details about the propagator at finite temperature are found in \cite{first}.

In the next section, the TFD formalism is used to calculate the differential cross section at finite temperature for Bhabha scattering in VSR.

\section{Differential cross section for Bhabha scattering in VSR at finite temperature}

In this section, the differential cross section for the Bhabha scattering in VSR at finite temperature is calculated. This scattering process corresponds to $e^-(p_1)e^+(p_2)\rightarrow e^-(q_1)e^+(q_2)$. The Feynman diagrams, that describe this scattering process are given in FIG. 1.
\begin{figure}[h]
\includegraphics[scale=0.7]{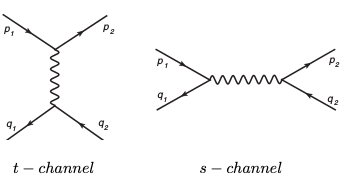}
\caption{Feynman diagrams to Bhabha scattering.}
\end{figure}

To calculate the differential cross section of Bhabha scattering in VSR, first the transition amplitude at finite temperature is determined. It is defined as
\bea
{\cal M}(\beta)=\langle f,\beta| \hat{S}^{(2)}| i,\beta\rangle,
\eea
with $\hat{S}^{(2)}$ being the second order term of the $\hat{S}$-matrix that is given as
\bea
\hat{S}&=&\sum_{n=0}^\infty\frac{(-i)^n}{n!}\int dx_1dx_2\cdots dx_n \mathbb{T} \left[ \hat{H}_{I}(x_1) \hat{H}_{I}(x_2)\cdots \hat{H}_{I}(x_n) \right],
\eea
where $\mathbb{T}$ is the time ordering operator and $\hat{H}_{I}(x)={H}_{I}(x)-\tilde{H}_{I}(x)$ describes the interaction. The thermal states are
\bea
| i,\beta\rangle&=& b_{p_1}^\dagger(\beta)d_{p_2}^\dagger(\beta)|0(\beta)\rangle, \nonumber\\
 \langle f,\beta|&=& \langle0(\beta)| b_{q_1}^\dagger(\beta)d_{q_2}^\dagger(\beta),
\eea
with $b_{p_j}^\dagger(\beta)$ and $d_{p_j}^\dagger(\beta)$ being creation operators.

Considering only the physical part, which is the non-tilde part, the transition amplitude becomes
\bea
{\cal M}(\beta)=-\frac{1}{2}\int d^4x\, d^4y\,\langle f,\beta|\bar{\psi}(x){\cal V}^\mu\psi(x)\bar{\psi}(y){\cal V}^\nu\psi(y)A_\mu(x)A_\nu(y)| i,\beta\rangle,
\eea
where ${\cal V}^\mu$ is the vertex given in eq. (\ref{vertex}).

Using the free field solution for the fermions field,
\bea
\psi(x)=\sum_r\int\frac{d^3p}{(2\pi)^{3/2}}\sqrt{\frac{m_e}{E_p}}\left[b_r(p)u_r(p)e^{-ipx}+d_r^\dagger(p)v_r(p)e^{ipx}\right]
\eea
with $u_r(p)$ and $v_r(p)$ being Dirac spinors, the transition amplitude for the t-channel and s-channel diagrams are calculated. 

The transition amplitude corresponding to the t-channel diagram is
\bea
{\cal M}(\beta)_t&=&-N_p\sum_r\int \frac{d^3p}{(2\pi)^{3/2}}\int d^4x\, d^4y\,(\mathsf{u}^2-\mathsf{v}^2)^2\bar{u}_r(p_2){\cal V}^\mu(p_1,p_2) u_r(p_1)\bar{v}_r(q_1){\cal V}^\nu(q_1,q_2) v_r(q_2)\nonumber\\
&\times&e^{i(p_2-p_1)x}e^{-i(q_1-q_2)y}\langle0(\beta)|\mathbb{T}[A_\mu(x)A_\nu(y)]|0(\beta)\rangle,
\eea
where
\bea
N_p\equiv \sqrt{\frac{m_e}{E_{p_1}}}\sqrt{\frac{m_e}{E_{p_2}}}\sqrt{\frac{m_e}{E_{q_1}}}\sqrt{\frac{m_e}{E_{q_2}}},
\eea
\bea
\int \frac{d^3p}{(2\pi)^{3/2}}\equiv \int \frac{d^3p_1}{(2\pi)^{3/2}}\int \frac{d^3p_2}{(2\pi)^{3/2}}\int \frac{d^3q_1}{(2\pi)^{3/2}}\int \frac{d^3q_2}{(2\pi)^{3/2}}
\eea
and the Bogoliubov transformations, eq. (\ref{f1}) and eq. (\ref{f2}), have been used. Using that $\mathsf{u}(\beta) =\cos \theta(\beta)$ and $\mathsf{v}(\beta) =\sin \theta(\beta)$ leads to $(\mathsf{u}^2-\mathsf{v}^2)^2= \tanh^2(\frac{\beta |q_0|}{2})$. Then
\bea
{\cal M}(\beta)_t&=&-N_p\sum_r\int \frac{d^3p}{(2\pi)^{3/2}}\int d^4x\, d^4y\,\tanh^2\left(\frac{\beta |q_0|}{2}\right)e^{i(p_2-p_1)x}e^{-i(q_1-q_2)y}\nonumber\\
&\times& \bar{u}_r(p_2){\cal V}^\mu(p_1,p_2) u_r(p_1)\bar{v}_r(q_1){\cal V}^\nu(q_1,q_2) v_r(q_2)\langle0(\beta)|\mathbb{T}[A_\mu(x)A_\nu(y)]|0(\beta)\rangle.
\eea

Considering that the photon propagator at finite temperature is defined as
\bea
\langle 0(\beta)|\mathbb{T}[A_\mu(x)A_\nu(y)]|0(\beta)\rangle=i\int \frac{d^4q}{(2\pi)^4}e^{-iq(x-y)}D_{\mu\nu}(q,\beta),
\eea
where $D_{\mu\nu}(q,\beta)$ is given in eq. (\ref{photon1}), the transition amplitude becomes
\bea
{\cal M}(\beta)_t&=&-iN_p\sum_r\int \frac{d^3p}{(2\pi)^{3/2}}\int d^4x\, d^4y\,\int \frac{d^4q}{(2\pi)^4}e^{-iq(x-y)}e^{i(p_2-p_1)x}e^{-i(q_1-q_2)y}\tanh^2\left(\frac{\beta |q_0|}{2}\right)\nonumber\\
&\times&\bar{u}_r(p_2){\cal V}^\mu(p_1,p_2) u_r(p_1)D_{\mu\nu}(q,\beta)\bar{v}_r(q_1){\cal V}^\nu(q_1,q_2) v_r(q_2).
\eea
Using the definition of the four-dimensional delta function,
\bea
\int d^4x\,d^4y\,e^{-ix(p_1-p_2+q)}e^{-iy(q_1-q_2-q)}=\delta^4(p_1-p_2+q)\delta^4(q_1-q_2-q),
\eea
and carrying out the q integral leads to
\bea
{\cal M}(\beta)_t&=&-iN_p\sum_r\int \frac{d^3p}{(2\pi)^{3/2}}\,\delta^4(p_1+q_1-p_2-q_2)\tanh^2\left(\frac{\beta |(p_1-p_2)_0|}{2}\right)\nonumber\\
&\times&\bar{u}_r(p_2){\cal V}^\mu(p_1,p_2) u_r(p_1)D_{\mu\nu}(p_1-p_2,\beta)\bar{v}_r(q_1){\cal V}^\nu(q_1,q_2) v_r(q_2),
\eea
where $|(p_1-p_2)_0|=|q_0|=\omega$ is the energy. The remaining delta function and the $p$ integral express overall four-momentum conservation. By convention, the final result is written as
\bea
{\cal M}(\beta)_t&=&-iN_p\sum_r \bar{u}_r(p_2){\cal V}^\mu(p_1,p_2) u_r(p_1)D_{\mu\nu}(p_1-p_2,\beta)\bar{v}_r(q_1){\cal V}^\nu(q_1,q_2) v_r(q_2)\nonumber\\
&\times&\tanh^2\left(\frac{\beta |(p_1-p_2)_0|}{2}\right).
\eea

The transition amplitude for the s-channel diagram is given as
\bea
{\cal M}(\beta)_s&=&iN_p\sum_r \bar{u}_r(p_2){\cal V}^\mu(p_1,-q_1) v_r(q_2)D_{\mu\nu}(p_1+q_1,\beta)\bar{v}_r(q_1){\cal V}^\nu(-p_2,q_2) u_r(p_1)\nonumber\\
&\times&\tanh^2\left(\frac{\beta |(p_1+q_1)_0|}{2}\right).
\eea

In order to calculate the differential cross section, the main quantity to be obtained is $|i{\cal M}(\beta)|^2$. Let's compute it by averaging over the spin of incoming and outgoing particles, then
\bea
\frac{1}{4}\sum_{spin}|i{\cal M}(\beta)|^2&=&\frac{1}{4}\sum_{spin}|{\cal M}(\beta)_t|^2+\frac{1}{4}\sum_{spin}|{\cal M}(\beta)_s|^2\nonumber\\
&+&\frac{1}{4}\sum_{spin}{\cal M}(\beta)_t^*{\cal M}(\beta)_s+\frac{1}{4}\sum_{spin}{\cal M}(\beta)_s^*{\cal M}(\beta)_t.
\eea

Using the completeness relations:
\bea
\sum_r u_r(p_1)\bar{u}_r(p_1)&=&\tilde{\slashed{p}}_1+m, \nonumber\\
\sum_r v_r(p_1)\bar{v}_r(p_1)&=&\tilde{\slashed{p}}_1-m,
\eea
where the wiggle momentum is $\tilde{p}_\mu=p_\mu-\frac{m^2 n_\mu}{2 (n\cdot p)}$, and the relation
\bea
\bar{v}(p_2)\gamma_\alpha u(p_1)\bar{u}(p_1)\gamma^\alpha v(p_2)=\mathrm{tr}\left[\gamma_\alpha u(p_1)\bar{u}(p_1)\gamma^\alpha  v(p_2)\bar{v}(p_2)\right].
\eea
In addition the center of mass (CM) frame is considered. The coordinates of CM are 
\bea
p_1&=&(E,\vec{p}),\quad\quad p_2=(E,-\vec{p}),\nonumber\\
q_1&=&(E,\vec{q})\quad\quad\mathrm{and}\quad\quad q_2=(E,-\vec{q}),
\eea
where $|\vec{p}|^2=|\vec{q}|^2=E^2$, $\vec{p}\cdot\vec{q}=E^2\cos\theta$ and $s=(2E)^2=E_{CM}^2$, we get $|(p_1-p_2)_0|=|(p_1+q_1)_0|=E_{CM}$. Thus, assuming that scattering process is calculated at the high energy limit, that corresponds to take $m_e^2=0$, the squared transition amplitudes become
\bea
\sum_{spin}|{\cal M}(\beta)_t|^2&=&\frac{e^4N_p^2}{16m_e^4}\tanh^4\left(\frac{\beta E_{CM}}{2}\right)\mathrm{tr}[{\cal V}_\mu(q_1,q_2)\gamma\cdot\tilde{q_2}{\cal V}_\alpha(q_1, q_2)\gamma\cdot\tilde{q_1}]\nonumber\\
&\times& \mathrm{tr}[{\cal V}^\mu(p_1,p_2)\gamma\cdot\tilde{p_1}{\cal V}^\alpha(p_1, p_2)\gamma\cdot\tilde{p_2}]\left[\frac{1}{(\tilde{p}_1-\tilde{p}_2)^4}+\Pi(\beta)\delta^2(\tilde{p}_1-\tilde{p}_2)\right],
\eea

\bea
\sum_{spin}|{\cal M}(\beta)_s|^2&=&\frac{e^4N_p^2}{16m_e^4}\tanh^4\left(\frac{\beta E_{CM}}{2}\right)\mathrm{tr}[{\cal V}_\mu(-p_2,q_2)\gamma\cdot\tilde{q_2}{\cal V}_\alpha(-p_2, q_2)(-\gamma\cdot\tilde{p_2})]\nonumber\\
&\times& \mathrm{tr}[{\cal V}^\mu(p_1,-q_1)\gamma\cdot\tilde{p_1}{\cal V}^\alpha(p_1, -q_1)(-\gamma\cdot\tilde{q_1})]\left[\frac{1}{(\tilde{p}_1+\tilde{q}_1)^4}+\Pi(\beta)\delta^2(\tilde{p}_1+\tilde{q}_1)\right],
\eea

\bea
\sum_{spin}{\cal M}(\beta)_s^*{\cal M}(\beta)_t&=&-\frac{e^4N_p^2}{16m_e^4}\mathrm{tr}[{\cal V}_\mu(p_1,-q_1)\gamma\cdot\tilde{p_2}{\cal V}^\alpha(p_1, p_2)\gamma\cdot\tilde{p_1}{\cal V}_\mu(-p_2,q_2)\gamma\cdot\tilde{q_1}{\cal V}_\alpha(q_1,q_2)\gamma\cdot\tilde{q_2}]\nonumber\\
&\times& \Bigl[\frac{1}{(\tilde{p}_1+\tilde{q}_1)^2(\tilde{p}_1-\tilde{p}_2)^2}+\frac{\Pi^{1/2}(\beta)i\delta(\tilde{p}_1+\tilde{q}_1)}{(\tilde{p}_1-\tilde{p}_2)^2}-\frac{\Pi^{1/2}(\beta)i\delta(\tilde{p}_1-\tilde{p}_2)}{(\tilde{p}_1+\tilde{q}_1)^2}\nonumber\\
&+&\Pi(\beta)\delta(\tilde{p}_1+\tilde{q}_1)\delta(\tilde{p}_1-\tilde{p}_2)\Bigl]\tanh^4\left(\frac{\beta E_{CM}}{2}\right)
\eea
and
\bea
\sum_{spin}{\cal M}(\beta)_t^*{\cal M}(\beta)_s&=&-\frac{e^4N_p^2}{16m_e^4}\mathrm{tr}[{\cal V}^\mu(p_1,p_2)(-\gamma\cdot\tilde{q_1}){\cal V}^\alpha(p_1, -q_1)\gamma\cdot\tilde{p_1}{\cal V}_\mu(q_1,q_2)(-\gamma\cdot\tilde{p_2})\nonumber\\
&\times&{\cal V}_\alpha(-p_2,q_2)\gamma\cdot\tilde{q_2}] \Bigl[\frac{1}{(\tilde{p}_1-\tilde{p}_2)^2(\tilde{p}_1+\tilde{q}_1)^2}+\frac{\Pi^{1/2}(\beta)i\delta(\tilde{p}_1-\tilde{p}_2)}{(\tilde{p}_1+\tilde{q}_1)^2}\nonumber\\
&-&\frac{\Pi^{1/2}(\beta)i\delta(\tilde{p}_1+\tilde{q}_1)}{(\tilde{p}_1-\tilde{p}_2)^2}+\Pi(\beta)\delta(\tilde{p}_1-\tilde{p}_2)\delta(\tilde{p}_1+\tilde{q}_1)\Bigl]\tanh^4\left(\frac{\beta E_{CM}}{2}\right),
\eea
where
\bea
\Pi(\beta)\equiv \frac{4\pi^2}{(e^{\beta E_{CM}}-1)^2}\left( \begin{array}{cc}1&e^{\beta E_{CM}/2}\\e^{\beta E_{CM}/2}&1\end{array} \right)^2.
\eea

Then the total square transition amplitude becomes
\bea
\frac{1}{4}\sum_{spin}|i{\cal M}(\beta)|^2&=&\frac{e^4}{64E^4}\tanh^4\left(\frac{\beta E_{CM}}{2}\right)\Biggl[\frac{{\cal J}_1}{(\tilde{p}_1-\tilde{p}_2)^4}+\frac{{\cal J}_2}{(\tilde{p}_1+\tilde{q}_1)^4}-\frac{{\cal J}_3}{(\tilde{p}_1-\tilde{p}_2)^2(\tilde{p}_1+\tilde{q}_1)^2}\nonumber\\
&+&\Pi(\beta)\left({\cal J}_1\delta^2(\tilde{p}_1-\tilde{p}_2)+{\cal J}_2\delta^2(\tilde{p}_1+\tilde{q}_1)-{\cal J}_3\delta(\tilde{p}_1+\tilde{q}_1)\delta(\tilde{p}_1-\tilde{p}_2)\right)\Biggl],
\eea
with
\bea
{\cal J}_1&=&\mathrm{tr}[{\cal V}_\mu(q_1,q_2)\gamma\cdot\tilde{q_2}{\cal V}_\alpha(q_1, q_2)\gamma\cdot\tilde{q_1}]\mathrm{tr}[{\cal V}^\mu(p_1,p_2)\gamma\cdot\tilde{p_1}{\cal V}^\alpha(p_1, p_2)\gamma\cdot\tilde{p_2}],\\
{\cal J}_2&=&\mathrm{tr}[{\cal V}_\mu(-p_2,q_2)\gamma\cdot\tilde{q_2}{\cal V}_\alpha(-p_2, q_2)(-\gamma\cdot\tilde{p_2})]\mathrm{tr}[{\cal V}^\mu(p_1,-q_1)\gamma\cdot\tilde{p_1}{\cal V}^\alpha(p_1, -q_1)(-\gamma\cdot\tilde{q_1})],\\
{\cal J}_3&=&\mathrm{tr}[{\cal V}_\mu(p_1,-q_1)\gamma\cdot\tilde{p_2}{\cal V}^\alpha(p_1, p_2)\gamma\cdot\tilde{p_1}{\cal V}_\mu(-p_2,q_2)\gamma\cdot\tilde{q_1}{\cal V}_\alpha(q_1,q_2)\gamma\cdot\tilde{q_2}]\nonumber\\
&+&\mathrm{tr}[{\cal V}^\mu(p_1,p_2)(-\gamma\cdot\tilde{q_1}){\cal V}^\alpha(p_1, -q_1)\gamma\cdot\tilde{p_1}{\cal V}_\mu(q_1,q_2)(-\gamma\cdot\tilde{p_2}){\cal V}_\alpha(-p_2,q_2)\gamma\cdot\tilde{q_2}].
\eea
Performing the trace that involve the product of up to eight gamma matrices, we get
\bea
{\cal J}_1&=&32(\tilde{q_2}\cdot\tilde{p_2})(\tilde{q_1}\cdot\tilde{p_1})+32(\tilde{p_1}\cdot\tilde{q_2})(\tilde{q_1}\cdot\tilde{p_2})\nonumber\\
&+&16m^2{\cal Q}_1\Bigl[(\tilde{q_2}\cdot\tilde{p_2})(n\cdot\tilde{q_1})(n\cdot\tilde{p_1})+(\tilde{q_2}\cdot\tilde{p_1})(n\cdot\tilde{q_1})(n\cdot\tilde{p_2})-2(\tilde{p_2}\cdot\tilde{p_1})(n\cdot\tilde{q_2})(n\cdot\tilde{q_1})\nonumber\\
&+&(\tilde{q_1}\cdot\tilde{p_2})(n\cdot\tilde{q_2})(n\cdot\tilde{p_1})+(\tilde{q_1}\cdot\tilde{p_1})(n\cdot\tilde{q_2})(n\cdot\tilde{p_2})-2(\tilde{q_2}\cdot\tilde{q_1})(n\cdot\tilde{p_1})(n\cdot\tilde{p_2})\Bigl],
\eea
\bea
{\cal J}_2&=&32(\tilde{q_2}\cdot\tilde{q_1})(\tilde{p_2}\cdot\tilde{p_1})+32(\tilde{p_1}\cdot\tilde{q_2})(\tilde{q_1}\cdot\tilde{p_2})\nonumber\\
&-&16m^2{\cal Q}_2\Bigl[(\tilde{q_2}\cdot\tilde{q_1})(n\cdot\tilde{p_2})(n\cdot\tilde{p_1})+(\tilde{q_2}\cdot\tilde{p_1})(n\cdot\tilde{q_1})(n\cdot\tilde{p_2})-2(\tilde{q_1}\cdot\tilde{p_1})(n\cdot\tilde{q_2})(n\cdot\tilde{p_2})\nonumber\\
&+&(\tilde{q_1}\cdot\tilde{p_2})(n\cdot\tilde{q_2})(n\cdot\tilde{p_1})+(\tilde{p_2}\cdot\tilde{p_1})(n\cdot\tilde{q_2})(n\cdot\tilde{q_1})-2(\tilde{q_2}\cdot\tilde{p_2})(n\cdot\tilde{q_1})(n\cdot\tilde{p_1})\Bigl]
\eea
and
\bea
{\cal J}_3&=&-64(\tilde{q_1}\cdot\tilde{p_2})(\tilde{q_2}\cdot\tilde{p_1})\nonumber\\
&+&16m^2({\cal Q}_1+{\cal Q}_2)\Bigl[-(n\cdot\tilde{p_2})(n\cdot\tilde{q_2})(\tilde{q_1}\cdot\tilde{p_1})+(n\cdot\tilde{p_1})(n\cdot\tilde{p_2})(\tilde{q_1}\cdot\tilde{q_2})\nonumber\\
&-&(n\cdot\tilde{q_1})(n\cdot\tilde{p_1})(\tilde{q_2}\cdot\tilde{p_2})+(n\cdot\tilde{q_1})(n\cdot\tilde{q_2})(\tilde{p_2}\cdot\tilde{p_1})\Bigl],
\eea
with
\bea
{\cal Q}_1&=&\frac{1}{(n\cdot\tilde{p_1})(n\cdot\tilde{p_2})}+\frac{1}{(n\cdot\tilde{q_1})(n\cdot\tilde{q_2})}\\
{\cal Q}_2&=&\frac{1}{(n\cdot\tilde{p_1})(n\cdot\tilde{q_1})}+\frac{1}{(n\cdot\tilde{p_2})(n\cdot\tilde{q_2})}.
\eea
Note that corrections due to the VSR non-local effects have been considered up to $m^2$ order.

Using these results and the CM coordinates, the differential cross section for Bhabha scattering in VSR at finite temperature, which is defined as
\bea
\left(\frac{d\sigma}{d\Omega}\right)_\beta=\frac{E^2}{16\pi^2}\frac{1}{4}\sum_{\mathrm{spins}}|i{\cal M(\beta)}|^2, \label{CS}
\eea
becomes,
\bea
\left(\frac{d\sigma}{d\Omega}\right)_\beta &=&\tanh^4\left(\frac{\beta E_{CM}}{2}\right)\Biggl[\frac{e^4(\cos(2\theta)+7)^2}{256\pi^2E_{CM}^2(\cos\theta-1)^2}\nonumber\\
&+&\chi^2\frac{e^4\left(642\cos\theta -80\cos 2\theta +125\cos 3\theta- 4\cos 4\theta +\cos 5\theta -172\right)}{8192\pi^2E^2(1-\cos\theta)^3}\Biggl]\nonumber\\
&+&32E^4\Pi(\beta)\tanh^4\left(\frac{\beta E_{CM}}{2}\right)\Bigl\{\left[\cos(2\theta)+3-8\chi^2\right]\delta^2(\tilde{p}_1-\tilde{p}_2)\nonumber\\
&+&\left[(\cos\theta+1)^2+4-8\chi^2(\cos\theta+1)\right]\delta^2(\tilde{p}_1+\tilde{q}_1)\nonumber\\
&-&4\cos^2\left(\theta/2\right)\left(\cos\theta+1+4\chi^2\right)\delta(\tilde{p}_1+\tilde{q}_1)\delta(\tilde{p}_1-\tilde{p}_2)\Bigl\}\label{res}
\eea
where $\chi=\frac{m}{E}$ is the parameter that controls the VSR effects. It is important to note that the photon propagator at finite temperature introduces product of delta functions with identical arguments \cite{Das1, Dolan, Das2, Das3}. This problem is avoided by working with the regularized form of delta-functions and their derivatives \cite{Van}:
\bea
2\pi i\delta^{(n)}(x)=\left(-\frac{1}{x+i\epsilon}\right)^{n+1}-\left(-\frac{1}{x-i\epsilon}\right)^{n+1}.
\eea

Results in eq. (\ref{res}), show that the cross section of Bhabha scattering has contributions due to VSR effects and due to finite temperature. In addition, the finite temperature effect modifies both results, that is,  the standard QED result and the VSR results. Furthermore, these results at finite temperatures are very important since the Lorentz violation is expected to appear at very high energy and very high temperature.

An important note, in the limit $T\rightarrow 0$ and $m^2\rightarrow 0$, we get
\bea
\left(\frac{d\sigma}{d\Omega}\right)_{QED} &=&\frac{e^4(\cos(2\theta)+7)^2}{256\pi^2E_{CM}^2(\cos\theta-1)^2}.
\eea
It is the usual QED cross section of the Bhabha scattering.

In addition, it is interesting to observe that, the Bhabha scattering is widely used to measure the luminosity. Concept of luminosity can be considered for colliding beams, for fixed targets and for Gaussian beams colliding head-on. In particle physics experiments the energy available for the production of
new effects is the most important parameter. Large amount of energy can only be provided with colliding beams where little or no energy is lost in the center of mass motion. This is specially important for rare events where little or very little energy is lost. A study of rare events with a small production cross section. Number of interactions lead to luminosity and thus provides number of events per second depends on the cross section and the luminosity.

The overall luminosity depends also on the factors like crossing angle, collision offset, hourglass effect, non-gaussian beam particles, non-zero dispersion at collision point and integrated luminosity. In addition it depends on optimization of integrated luminosity and space and time structure of luminosity. There additional effects that play a role in absolute measurement of profile measurement and beam displacement.

For collision of two particles with mass $m_1$ and $m_2$, the total center of mass energy may be expressed as $(p_1+p_2)^2=(E_1+E_2)^2-(p_1+p_2)^2$. This leads to different center of mass energy that will be for energy as collider or for a fixed target. This suggests that for colliding beams it is necessary to have high center of mass energy.
   
Similarly the fixed target and colliding beam luminosity are quite distinct. In this case the role of beams with distinct luminosity the overall impact is quite different. It is important to mention that there are additional complications for real machines like crossing angle, hourglass effect and crossing angles.

All this and other factors like integrated luminosity, space and time structure of luminosity lead to the overall impact on the over structure of the system.  These factors play an important role in the study of the Bhabha scattering in very special relativity at finite temperature. These ideas are important to fully understand the Lorentz-violating operators at finite temperature on the cross section. Furthermore, for very small scattering angles ($\theta\ll 1$), the cross section of Bhabha scattering is very big making this scattering a well suited process for luminosity measurements. Therefore, constraints on Lorentz-violating parameter can be obtained if the precision of the measurements of luminosity for very small angle  will improve significantly. Although the study developed here is completely theoretical, our results show that temperature effects and experimental data for very small angle may contribute to a new class of constraints on VSR parameter.

\section{Conclusion} 

The Very Special Relativity has been considered. In this theory laws of physics are not invariant under the whole Poincaré group but rather under subgroups of the Poincaré group. In this context a modified gauge symmetry is present admitting a variety of new gauge invariant interactions. The Poincaré subgroup is locally invariant under the symmetries $SIM(2)$ or $HOM(2)$ and Lorentz. However, non-local terms are constructed. Then these non-local terms are $SIM(2)$ or $HOM(2)$ invariant, but break the Lorentz symmetry. Our main objective is to study the Bhabha scattering in VSR framework. The corrections due to VSR are obtained. In addition, corrections due to finite temperature in Bhabha scattering are calculated. The TFD formalism is used to introduce finite temperature. Our results show that the usual differential cross section of QED is changed due to VSR and temperature effects. Although the high energy experiments at low temperatures, it is very interesting to investigate scattering process at very high temperatures. This gives us a good estimate of the importance of the Lorentz-violating operators at finite temperature on the cross section.

\section*{Acknowledgments}

This work by A. F. S. is supported by CNPq projects 308611/2017-9 and 430194/2018-8.

\end{document}